\documentclass[12pt]{article}

\usepackage{graphicx}

\newcommand{\Z}{{\sf Z \!\!\! Z}}

\newcommand{\1}{{\sf 1 \!\! 1}}
\newcommand{\Psibar}{\bar{\Psi}}
\makeatletter
\@addtoreset{equation}{section}
\makeatother

\title{QCD at Fixed Topology}

\author{R.Brower$^{{\rm a,b}}$, S. Chandrasekharan$^{{\rm c,d}}$, 
J. W. Negele$^{{\rm b}}$ and U.-J. Wiese$^{{\rm b,d}}$ \\ \\
$^{{\rm a}}$ Department of Physics, Boston University \\
Boston, Massachusetts 02215, U.S.A. \\ \\
$^{{\rm b}}$ Center for Theoretical Physics, \\
Laboratory for Nuclear Science, and Department of Physics \\
Massachusetts Institute of Technology (MIT) \\
Cambridge, Massachusetts 02139, U.S.A. \\ \\
$^{{\rm c}}$ Department of Physics, Duke University \\
Durham, North Carolina 27708, U.S.A. \\ \\
$^{{\rm d}}$ Institute for Theoretical Physics, Bern University \\
3012 Bern, Switzerland \\ \\
DUKE-TH-02-229, MIT-CTP-3201 \\ \\}
 
\begin{document}

\maketitle
\vspace{-.1in}
\begin{abstract} \normalsize
Since present Monte Carlo algorithms for lattice QCD may become trapped
in a fixed topological charge sector, it is important to understand the effect 
of calculating at fixed topology. In this work, we show that although the 
restriction to a fixed topological sector becomes irrelevant in the infinite 
volume limit, it gives rise to characteristic finite size effects due to 
contributions from all $\theta$-vacua. We calculate these effects and show how 
to extract physical results from numerical data obtained at fixed topology.

\end{abstract}
 
\maketitle
 
\newpage

\section{Introduction}

The numerical solution of lattice QCD is a notoriously hard problem. In
particular, in order to obtain physical results, the numerical data must be 
extrapolated to the chiral, continuum, and infinite volume limits. Fortunately,
the dependence of physical quantities on the quark masses, the lattice spacing,
or the finite box size can often be understood analytically, and formulae can 
be derived that allow us to extrapolate reliably to the physical limit. Here we
derive such formulae to correct for finite size effects that are due to fixed 
topology.

The configuration space of QCD decomposes into topological sectors. With 
boundary conditions that are periodic up to gauge transformations the 
topological charge $Q$ is an integer. In the continuum theory different
topological sectors are separated by infinite action barriers. In a lattice 
theory the action barriers depend on the choice of the lattice action and the 
lattice definition of the topological charge. Although on the lattice the 
action barriers are typically finite, they may cause problems in the numerical 
sampling of the various $Q$ sectors. In particular, a standard algorithm
(e.g. the hybrid Monte Carlo algorithm) that changes the gauge field 
configuration in small steps may get stuck in a fixed topological charge 
sector because it cannot overcome the action barriers. The question arises how 
one can extract meaningful physical information from numerical calculations 
that get stuck in a given topological sector. Locality and cluster 
decomposition 
properties suggest that the restriction to fixed topology becomes irrelevant in
the infinite volume limit. Still, in order to extrapolate reliably to that 
limit, one must understand the corresponding finite size effects.

The Atiyah-Singer index theorem, $Q = n_L - n_R$, connects the physics of light
quarks to the topology of the gluon field: the difference of the number of 
left- and right-handed zero modes of the Dirac operator, $n_L$ and $n_R$, is 
equal to the 
topological charge. Consequently, when one quark is exactly massless, the zero 
modes of the Dirac operator eliminate all nontrivial topological charge sectors
(with $Q \neq 0$) from the QCD partition function. Hence, working at fixed 
$Q = 0$ is then correct even in a finite volume. However, even in that case 
various physical observables (e.g. the chiral condensate) get contributions 
from 
$Q \neq 0$ sectors, and one still needs to sample several topological sectors 
in order to compute those quantities. In the presence of a massless quark, the 
complex phase of the Boltzmann factor $\exp(i \theta Q)$ for a $\theta$-vacuum 
is trivial and the physics is independent of $\theta$. When all quarks are 
massive, there are nontrivial $\theta$-vacua effects and all topological 
sectors contribute to the partition function. In small space-time volumes with 
$\beta V \langle \Psibar \Psi \rangle m \ll 1$ (where $\beta$ is the inverse 
temperature, $V$ is the spatial volume, $\langle \Psibar \Psi \rangle$ is the 
chiral condensate, and $m$ is the quark mass) the partition function is still 
dominated by the $Q = 0$ contribution. On the other hand, in large volumes with
$\beta V \langle \Psibar \Psi \rangle m \gg 1$ one must sample a large number 
of topological charge sectors. 

When standard lattice fermion actions are used, the relation between the
topology of the gluon field and the zero modes of the Dirac operator is
obscured by lattice artifacts. In that case, the effects that we discuss here
are difficult to observe in a numerical computation. However, when one uses
a fermion action that obeys the Ginsparg-Wilson relation \cite{Gin82}, the 
chiral properties of continuum fermions persist on the lattice. Our results ---
which we derive for the continuum theory --- directly apply to this type of
lattice fermions. 

The rest of the paper is organized as follows. In section 2 the basic physics 
of $\theta$-vacua and the topological charge is reviewed, and in section 3 a 
simple formula for the fixed topology finite size effect on particle masses is 
derived. In section 4 chiral perturbation theory is used to study the 
$\theta$-dependence of the vacuum energy and the pion mass. In section 5, we 
explore the large $N_c$ limit in chiral perturbation theory. Since current 
lattice calculations are performed far from the chiral limit, in section 6 the 
$Q$-dependence of the $\eta'$-mass arising in an instanton gas model is 
estimated and compared with lattice results. Finally, section 7 contains the 
conclusions.

\section{Topological Charge Sectors and $\theta$-Vacua}

The partition function of QCD in a $\theta$-vacuum is given by
\begin{equation}
\label{thetaqcd}
Z(\theta) = \int {\cal D}A {\cal D}\Psibar {\cal D}\Psi \ 
\exp(- S[A,\Psibar,\Psi]) \exp(- i \theta Q[A]),
\end{equation}
where
\begin{equation}
Q[A] = \frac{1}{32 \pi^2} \int d^4x \ \epsilon_{\mu\nu\rho\sigma}
\mbox{Tr} F_{\mu\nu} F_{\rho\sigma}
\end{equation}
is the topological charge of the gluon field $A_\mu(x) = i A_\mu^a(x)
\lambda^a$ with field strength
\begin{equation}
F_{\mu\nu}(x) = \partial_\mu A_\nu(x) - \partial_\nu A_\mu(x) + 
[A_\mu(x),A_\nu(x)],
\end{equation}
and $S[A,\Psibar,\Psi]$ is the Euclidean QCD action, which is invariant under
gauge transformations
\begin{equation}
A'_\mu(x) = g(x)(A_\mu(x) + \partial_\mu)g(x)^\dagger, \
\Psi'(x) = g(x) \Psi(x), \ \Psibar'(x) = \Psibar(x) g(x)^\dagger.
\end{equation}
We consider the theory in 
a finite spatial volume $V = L_x L_y L_z$ with periodic boundary conditions at 
temperature $T$, i.e. with finite Euclidean time extent $L_t = \beta = 1/T$. In
order to allow for nontrivial topology, it is necessary to impose periodicity 
only for gauge invariant quantities. Thus, $A_\mu(x)$, $\Psi(x)$ and 
$\Psibar(x)$ are periodic only up to gauge transformations 
\begin{eqnarray}
&&A_\mu(x + L_\mu \hat\mu) = g_\mu(x)(A_\mu(x) + \partial_\mu)g_\mu(x)^\dagger,
\nonumber \\
&&\Psi(x + L_\mu \hat\mu) = g_\mu(x) \Psi(x), \
\Psibar(x + L_\mu \hat\mu) = \Psibar(x) g_\mu(x)^\dagger.
\end{eqnarray}
Here $\hat\mu$ is the unit-vector in the $\mu$-direction and 
the $g_\mu(x)$ are a set of transition functions that define the boundary
condition. As first realized by 't Hooft \cite{tHo79}, for an $SU(N_c)$ 
Yang-Mills theory these boundary conditions fall into gauge equivalence classes
labeled by a twist tensor that takes values in the center of the gauge group. 
In general, for a nontrivial twist tensor, the topological charge can take 
fractional values that are multiples of $1/N_c$ \cite{vBa82}. In the presence 
of matter fields that transform nontrivially under the center (e.g. quarks in 
the fundamental representation of $SU(3)$), single valuedness of the matter 
field implies that the twist tensor is trivial. Hence, in the QCD case the 
topological charge is an integer and not a fraction. It is interesting to note
that in supersymmetric Yang-Mills theory the gluinos, which transform under 
the adjoint representation, leave the center symmetry unbroken. Consequently, 
one then expects fractional topological charges. The role of the topological 
charge in QCD at finite volume has been worked out in great detail by 
Leutwyler and Smilga \cite{Leu92}. Although eq.(\ref{thetaqcd}) is quite 
formal, it can be given a concrete meaning on a space-time lattice using, for 
example, Ginsparg-Wilson fermions. Below we will ignore this subtlety and 
perform formal manipulations on eq.(\ref{thetaqcd}), assuming that all of 
these can be given a concrete realization on the lattice.

If we define $H$ to be the QCD Hamiltonian with the energy eigenstates 
$|n,\theta \rangle$ in a given $\theta$-vacuum, then we can write
\begin{equation}
Z(\theta) = \sum_n \langle n,\theta|\exp(- \beta H)|n,\theta\rangle =
\sum_n \exp(- \beta E_n(\theta)).
\end{equation}
For sufficiently large spatial volumes, the $\theta$-vacuum energy is given by
$E_0(\theta) = V e_0(\theta)$, where $e_0(\theta)$ is the vacuum energy 
density. For small values of $\theta$ it is given by
\begin{equation}
\label{e0chit}
e_0(\theta) = \frac{1}{2} \chi_t \theta^2 + \gamma \theta^4 + ...,
\end{equation}
where
\begin{equation}
\chi_t = \frac{\langle Q^2 \rangle}{\beta V},
\end{equation}
is the topological susceptibility (at $\theta = 0$). 
For $\beta [E_1(\theta)-E_0(\theta)] \gg 1$ the vacuum state dominates and 
we get 
\begin{equation}
\label{Ztheta}
Z(\theta) = \exp(- \beta V e_0(\theta)) = \exp(- \frac{\beta V \chi_t}{2} 
\theta^2)[1 + {\cal O}(\gamma \theta^4)].
\end{equation}
For small values of $\theta$ and for $M_\pi L \gg 1$
(where $L = L_x = L_y = L_z$) the first excited state is the pion, i.e.
$E_1-E_0 = M_\pi$. On the other hand, in the limit of massless quarks chiral 
symmetry is restored in a finite volume and a rotor spectrum of states exists 
above the vacuum \cite{Leu87}. In that case, the energy of the first excited 
state is given by
\begin{equation}
E_1(0) - E_0(0) = \frac{N_f^2 - 1}{N_f F_\pi^2 V},
\end{equation}
where $N_f$ is the number of quark flavors and $F_\pi$ is the pion decay
constant. In other words, the energy gap is inversely proportional to the 
spatial volume. As the quark mass increases, this state evolves into the state 
of a pion at rest. For $\theta = \pi$ (depending on the quark masses and $N_f$)
the CP symmetry breaks spontaneously \cite{Wit80}. In that case, there are two 
degenerate $\theta$-vacua in the infinite volume. In a finite volume, due to 
tunneling, there is an exponentially small energy gap
\begin{equation}
E_1(\pi) - E_0(\pi) = A \exp(- \alpha V).
\end{equation}
Here $\alpha$ is the tension of a 3-dimensional interface separating Euclidean
space-time regions that are in distinct CP phases. 

The QCD partition function at fixed topological charge $Q$ is given by
\begin{equation}
Z_Q = \int {\cal D}A {\cal D}\Psibar {\cal D}\Psi \ \delta_{Q,Q[A]}
\exp(- S[A,\Psibar,\Psi]).
\end{equation}
Writing the Kronecker $\delta$-function as
\begin{equation}
\delta_{Q,Q[A]} = \frac{1}{2 \pi} \int_{-\pi}^\pi d\theta \ 
\exp[i \theta (Q - Q[A])],
\end{equation}
the fixed $Q$ partition function turns into an integral over all 
$\theta$ values
\begin{eqnarray}
Z_Q&=&\frac{1}{2 \pi} \int_{-\pi}^\pi d\theta \ Z(\theta) \exp(i \theta Q)
\nonumber \\
&=&\frac{1}{2 \pi} \int_{-\pi}^\pi d\theta \
\sum_n \exp(- \beta E_n(\theta)) \exp(i \theta Q).
\end{eqnarray}
At sufficiently low temperatures ($\beta E_1(0) \gg 1$) and sufficiently small
topological charges (i.e. $|Q|/\sqrt{\beta V}$ remains finite) the partition 
function reduces to
\begin{equation}
\label{partition}
Z_Q =\frac{1}{2 \pi} \int_{-\pi}^\pi d\theta \
\exp(- \beta V e_0(\theta)) \exp(i \theta Q).
\end{equation}
This equation is the starting point to derive the effects that we are 
interested in. For large space-time volumes $\beta V$, the fixed $Q$ 
partition function is dominated by the $\theta$-vacua with a small energy 
density. Since the $\theta = 0$ vacuum has the smallest energy, we see that the
physics at large $\beta V$ is dominated by $\theta = 0$. Hence, we can use a 
saddle point approximation to obtain a formula for $Z_Q$. We assume a large
$\beta V$ keeping $Q/\sqrt{\beta V}$ finite and fixed. Using eq.(\ref{e0chit}) 
one obtains
\begin{eqnarray}
\label{ZQ}
Z_Q&=&\sqrt{\frac{2 \pi}{\beta V \chi_t}}
\exp\left[- \frac{Q^2}{2 \beta V \chi_t}\right]
\left\{1 + {\cal O}(\frac{\gamma}{\beta V})\right\} \nonumber \\
&=&\sqrt{\frac{2 \pi}{\langle Q^2 \rangle}}
\exp\left[- \frac{Q^2}{2 \langle Q^2 \rangle}\right]
\left\{1 + {\cal O}(\frac{\gamma}{\beta V})\right\}.
\end{eqnarray}

\section{Effective Mass Determination at fixed $Q$}

It is possible to determine the hadron spectrum from a Monte Carlo simulation 
even if it gets stuck in a fixed topological sector. Typically hadron masses 
are extracted from 2-point correlation functions of operators with the 
appropriate quantum numbers. Higher $n$-point functions are important in the
determination of other physical quantities like matrix elements and coupling
constants. In a $\theta$-vacuum a general $n$-point correlation function of 
operators ${\cal O}_i$ takes the form
\begin{eqnarray}
G(\theta)&=&\langle {\cal O}_1 {\cal O}_2 ... {\cal O}_n \rangle_\theta
\nonumber \\
&=&\frac{1}{Z(\theta)}  \int {\cal D}A {\cal D}\Psibar {\cal D}\Psi \ 
{\cal O}_1 {\cal O}_2 ... {\cal O}_n \exp(- S[A,\Psibar,\Psi]) 
\exp(- i \theta Q[A]).
\end{eqnarray}
Hence, in a sector of fixed topological charge one obtains
\begin{equation}
G_Q = \langle {\cal O}_1 {\cal O}_2 ... {\cal O}_n \rangle_Q =
\frac{1}{Z_Q} \frac{1}{2 \pi} \int_{- \pi}^\pi d\theta \
Z(\theta) G(\theta) \exp(i \theta Q).
\end{equation}
Again, assuming small enough temperatures ($\beta E_1(0) \gg 1$) and 
sufficiently small topological charges (i.e. a finite $|Q|/\sqrt{\beta V}$), 
$Z_Q$ is given by eq.(\ref{ZQ}) and $Z(\theta)$ is given by eq.(\ref{Ztheta}).
Assuming that $\langle {\cal O}_1 {\cal O}_2 ... {\cal O}_n \rangle_\theta$ is
smooth over the range $\theta^2 < 1/\beta V \chi_t = 1/\langle Q^2 \rangle$, 
one can again use the saddle point method to evaluate the integral over 
$\theta$. Then one obtains
\begin{equation}
\label{Greenf}
G_Q = G(\theta_s) + \frac{1}{2 \beta V \chi_t} G''(\theta_s) + ...,
\end{equation}
where a prime denotes a derivative with respect to $\theta$. The (purely 
imaginary) value of $\theta$ at the saddle point is given by
\begin{equation}
\theta_s = i \frac{Q}{\beta V \chi_t} = i \frac{Q}{\langle Q^2 \rangle}.
\end{equation}

Let us now consider a 2-point correlation function
\begin{eqnarray}
&&\langle {\cal O}(t_1) {\cal O}(t_2) \rangle_\theta \nonumber \\
&&=\frac{1}{Z(\theta)}  
\int {\cal D}A {\cal D}\Psibar {\cal D}\Psi \ {\cal O}(t_1) {\cal O}(t_2) 
\exp(- S[A,\Psibar,\Psi]) \exp(- i \theta Q[A]) \nonumber \\ 
&&=\frac{1}{Z(\theta)}  
\sum_n \langle n,\theta|{\cal O} \exp(- H (t_1-t_2)) {\cal O} 
\exp(- H (\beta-t_1+t_2))|n,\theta\rangle \nonumber \\
&&=\frac{1}{Z(\theta)} \sum_{m,n} \exp(- \beta E_m(\theta))
|\langle m,\theta|{\cal O}|n,\theta \rangle|^2 
\exp[- (E_n(\theta) - E_m(\theta))(t_1-t_2)], \nonumber \\ \,
\end{eqnarray}
relevant for determining particle masses. Here the operator ${\cal O}$ has the
appropriate quantum numbers in order to create the particle of interest at
rest from the $\theta = 0$ vacuum. The $\theta$-dependent particle mass is then
given by $M(\theta) = E_1(\theta) - E_0(\theta)$. For large time separations 
$(t_1-t_2)$ and at sufficiently low temperatures (large $\beta$) the 
correlation function reduces to
\begin{equation}
\label{correlation}
\langle {\cal O}(t_1) {\cal O}(t_2) \rangle_\theta =
|\langle 0,\theta|{\cal O}|1,\theta \rangle|^2 \exp[- M(\theta)(t_1-t_2)].
\end{equation}
Using eq.(\ref{Greenf}) and assuming
\begin{equation}
M(\theta) = M(0) + \frac{1}{2} M''(0) \theta^2 + ...,
\end{equation}
one obtains
\begin{equation}
\langle {\cal O}(t_1) {\cal O}(t_2) \rangle_Q \sim A_Q \exp[- M_Q(t_1-t_2)],
\end{equation}
where $A_Q$ is a constant and
\begin{eqnarray}
\label{massQ}
M_Q &=& M(0) + \frac{1}{2} M''(0) \frac{1}{\beta V \chi_t}
(1 - \frac{Q^2}{\beta V \chi_t})+ ... \nonumber \\
&=& M(0) + \frac{1}{2} M''(0) \frac{1}{\langle Q^2 \rangle}
(1 - \frac{Q^2}{\langle Q^2 \rangle})+ ....
\end{eqnarray}
It should be noted that the concept of an effective ``mass'' $M_Q$ (which is
independent of $t_1 - t_2)$) in a fixed topological sector makes sense only to
the quadratic order of the saddle point expansion considered here. Higher order
terms lead to time-dependent ``masses''. This is not surprising because, due to
the topological constraint, QCD at fixed $Q$ is a nonlocal theory that does not
even have a Hamiltonian. We note again that we take the large
$\beta V$ limit keeping $Q/\sqrt{\beta V}$ fixed. Hence, the term
$Q^2/\langle Q^2 \rangle$ is of order 1. When $Q$ itself is fixed this
term becomes of order $1/\beta V$. In this case, there are other terms of the
same order that we have not calculated. Hence, in order to be consistent, this 
term should then be dropped.

The above discussion shows that if a calculation performed in a space-time 
volume $\beta V$ gets stuck in a fixed $Q$ sector, one would extract a mass 
$M_Q$ that deviates from the infinite volume result by a term proportional to 
$1/(\beta V)$. This term decreases rapidly as one increases the space-time 
volume. Of course, from a simulation that is stuck in a fixed topological 
charge sector, one will not be able to extract $\langle Q^2 \rangle$, because 
it will be impossible to determine the relative weight of different sectors. 
Still, one should be able to perform separate computations for the various 
topological sectors, simply by starting the calculation in a given $Q$ sector. 
Then, in principle,  one can measure $M_Q$ in the various sectors and at
various space-time volumes and thereby extract $M(0)$ , $M''(0)$ and 
$\langle Q^2 \rangle$ using eq.(\ref{massQ}). It is interesting to note that 
when $M_Q$ is averaged over $Q$ with the distribution 
$\exp(- Q^2/2 \langle Q^2 \rangle)$ one obtains the mass $M(0)$ in the 
$\theta = 0$ vacuum.

Let us briefly consider the extreme case of zero quark mass. Then the pion is 
massless and the box size is always small compared to the infinite Goldstone 
boson correlation length. For $m = 0$ there are no $\theta$-effects --- 
$Z(\theta)$ is independent of $\theta$ --- and the fixed $Q$ partition function
$Z_Q$ is non-zero only for $Q = 0$. Hence, in this case, it is correct to work 
at fixed topological charge $Q = 0$, unless one wants to measure observables
like the chiral condensate that receive contributions from $Q \neq 0$ sectors.

\section{$\theta$-Dependence of Physical Quantities}

Based on the previous discussion, it is clear that in a large space-time volume
the effects of fixed topology can be determined from the $\theta$-dependence of
certain physical quantities. For example, the partition function $Z_Q$ at fixed
topological charge $Q$ can be determined from the $\theta$-dependence of the 
vacuum energy $e_0(\theta)$ using eq.(\ref{partition}) and the effective mass 
of a hadron $M_Q$ can be obtained from $M(\theta)$ using eq.(\ref{massQ}).

Until now we have written the QCD action in a $\theta$-vacuum as 
$S[A,\Psi,\Psibar] + i \theta Q[A]$.
It is well known that, using an anomalous $U(1)_A$ chiral transformation, the
$\theta$-dependence can be moved into the quark mass matrix 
${\cal M} = \mbox{diag}(m_u,m_d,...,m_{N_f})$ (for $N_f$ flavors), i.e.
\begin{eqnarray}
S_\theta[A,\Psi,\Psibar]
&=&\frac{1}{g^2} \int d^4x \ \mbox{Tr} F_{\mu\nu}F_{\mu\nu} \nonumber \\
&+&\int d^4x \ \{\Psibar_L [\gamma_\mu (A_\mu + \partial_\mu)] \Psi_L + 
\Psibar_R [\gamma_\mu (A_\mu + \partial_\mu)] \Psi_R \} \nonumber \\ 
&+&\int d^4x \ \{\Psibar_L {\cal M} e^{i \theta/N_f} \Psi_R + 
\Psibar_R {\cal M}^\dagger e^{- i \theta/N_f} \Psi_L \}.
\end{eqnarray}
Thus, $\theta$ appears simply as the complex phase of the determinant of
the mass matrix

Although $\theta$ enters the QCD action in a simple way, in general its effects
on physical quantities are difficult to determine. However, for small quark
masses chiral perturbation theory allows us to understand such effects. It is 
well known that the low-energy physics of Goldstone bosons in QCD can be 
described systematically using an effective theory. In the case of $N_f$ light 
quarks to lowest order the action of the Goldstone boson field $U \in SU(N_f)$ 
takes the form
\begin{equation}
\label{Chact1}
S[U] = \int d^4x \ 
\{\frac{F_\pi^2}{4} \mbox{Tr}[\partial_\mu U^\dagger \partial_\mu U] -
\frac{\langle \Psibar \Psi \rangle}{2 N_f} 
\mbox{Tr}[{\cal M} e^{i \theta/N_f} U^\dagger + 
{\cal M}^\dagger e^{- i \theta/N_f} U]\},
\end{equation}
where $F_\pi$ is the pion decay constant. The term proportional to $F_\pi^2$ is
invariant under chiral $SU(N_f)_L \otimes SU(N_f)_R$ transformations
\begin{equation}
U'(x) = L U(x) R^\dagger,
\end{equation}
(with $L \in SU(N_f)_L$ and $R \in SU(N_f)_R$) while the mass term explicitly 
breaks chiral symmetry.

First, we consider the case of $N_f$ degenerate flavors of mass $m$. Then the 
mass matrix takes the form ${\cal M} = m \1$.
In this case, the chiral symmetry is explicitly broken down to the vector
subgroup $SU(N_f)_{L=R}$. When the vacuum angle changes by $2 \pi n$,
$\exp(i \theta/N_f)$ changes by an element $z = \exp(2 \pi n i/N_f) \in 
\Z(N_f)$ --- the center of the flavor group $SU(N_f)$. Although the action 
itself is not $2 \pi$-periodic in $\theta$, the corresponding partition 
function
\begin{equation}
Z(\theta) = \int {\cal D}U \ \exp(- S[U])
\end{equation}
is, because the shift $\exp(i \theta'/N_F) = z \exp(i \theta/N_f)$ can be
compensated by a redefinition of the field $U' = z U$. For $\theta \in 
[-\pi,\pi]$, the effective action is minimized by constant field configurations
$U = \1$. The value of the action at the minimum 
\begin{equation}
S[\1] = - \beta V \langle \Psibar \Psi \rangle m \cos(\theta/N_f)
\end{equation}
determines the $\theta$-vacuum energy density (normalized to zero at 
$\theta = 0$) as
\begin{equation}
\label{thetadensity}
e_0(\theta) = \langle \Psibar \Psi \rangle m (1 - \cos(\theta/N_f)).
\end{equation}
This expression is valid for $\theta \in [-2 \pi/N_f,2 \pi/N_f]$. When extended
periodically to other $\theta$-values, it has cusps at $\theta = 
4 \pi(n + \frac{1}{2})/N_f$ indicating a phase transition in the vacuum angle
\cite{Wit80}. Comparing this result with eq.(\ref{e0chit}), we identify the 
topological susceptibility as
\begin{equation}
\chi_t = \frac{\langle \Psibar \Psi \rangle m}{N_f^2}.
\end{equation}

In order to determine the $\theta$-dependence of the Goldstone boson mass, we 
expand $U = \exp(i \pi^a \lambda^a/F_\pi)$ in powers of $\pi^a$ and read off 
$M_\pi^2(\theta)$ as the coefficient of the $\pi^a \pi^a$ term. This yields
\begin{equation}
\label{thetamass}
M_\pi^2(\theta) = M_\pi^2(0) \cos(\theta/N_f),
\end{equation}
where
\begin{equation}
\label{pionmass}
M_\pi^2(0) = \frac{2 m \langle\Psibar \Psi \rangle}{N_f F_\pi^2},
\end{equation}
is the Goldstone boson mass squared at $\theta = 0$.

Using leading order baryon chiral perturbation theory \cite{Gas88} the quark 
mass dependence of the nucleon mass (at $\theta = 0$) is given by
\begin{equation}
M_N(0) = M_N^0[1 + c \frac{M^2_\pi}{F_\pi^2}] - 
\frac{3 g_A^2}{32 \pi} \frac{M_\pi^3}{F_\pi^2},
\end{equation}
where $M_N^0$ and $g_A$ are the nucleon mass and the neutron decay constant in 
the chiral limit, and $c$ is a known combination of low-energy constants. At 
this order in chiral perturbation theory the $\theta$-dependence enters only
through the pion mass of eq.(\ref{thetamass}) such that
\begin{equation}
M_N(0) = M_N^0[1 + c \frac{M^2_\pi \cos(\theta/N_f)}{F_\pi^2}] - 
\frac{3 g_A^2}{32 \pi} \frac{M_\pi^3 \cos^{3/2}(\theta/N_f)}{F_\pi^2},
\end{equation}

Let us now consider the case of two non-degenerate flavors with masses $m_u$
and $m_d$. Then the mass matrix takes the form ${\cal M} = 
\mbox{diag}(m_u,m_d)$, and chiral symmetry is explicitly broken down to 
$U(1)_{L=R}$. Now the action is minimized for 
$U = \mbox{diag}(\exp(i \varphi),\exp(- i \varphi))$ where
\begin{equation}
\tan\varphi = \frac{m_u - m_d}{m_u + m_d} \tan\frac{\theta}{2},
\end{equation}
and one obtains
\begin{equation}
e_0(\theta) = \langle \Psibar \Psi \rangle \frac{m_u + m_d}{2} 
\left[1 - \cos\left(\frac{\theta}{2}\right) 
\sqrt{1 + \frac{(m_u - m_d)^2}{(m_u + m_d)^2} 
\tan^2\frac{\theta}{2}}\ \right].
\end{equation}
The pion mass is given by
\begin{equation}
M_\pi^2(\theta) = M_\pi^2(0) \cos\left(\frac{\theta}{2}\right)
\sqrt{1 + \frac{(m_u - m_d)^2}{(m_u + m_d)^2} \tan^2\frac{\theta}{2}},
\end{equation}
where now
\begin{equation}
M_\pi^2(0) = \frac{\langle \Psibar \Psi \rangle (m_u + m_d)}{2 F_\pi^2}.
\end{equation}

\section{$\theta$-Dependence at large $N_c$}

The flavor-singlet pseudoscalar particle $\eta'$ is special in QCD. Due to the 
anomaly it is heavy (and thus not a Goldstone boson) and hence its physics 
cannot be understood using chiral perturbation theory. However, in the large 
$N_c$ limit the anomaly disappears and the $\eta'$ becomes a Goldstone boson
which can now be studied in chiral perturbation theory. The low-energy 
effective action of eq.(\ref{Chact1}) is modified to \cite{Wit80}
\begin{eqnarray}
\label{Chact2}
S[\tilde U]&=&\int d^4x \ \{\frac{F_\pi^2}{4} 
\mbox{Tr}[\partial_\mu \tilde U^\dagger \partial_\mu \tilde U] -
\frac{\langle \Psibar \Psi \rangle}{2 N_f} 
\mbox{Tr}[{\cal M} e^{i \theta/N_f} \tilde U^\dagger + 
{\cal M}^\dagger e^{- i \theta/N_f} \tilde U]\} \nonumber \\
&+&\frac{1}{2} \chi_t^0 (i \log \mbox{det} \tilde U)^2.
\end{eqnarray}
Here $\tilde U$ is a $U(N_f)$ matrix with complex determinant 
$\exp(i \sqrt{2 N_f} \eta'/F_\pi)$. The last term in eq.(\ref{Chact2}) gives
rise to the $\eta'$ mass
\begin{equation}
M_{\eta'}^2 = \frac{2 N_f \chi_t^0}{F_\pi^2}.
\end{equation}
At large $N_c$ the pion decay constant goes as $F_\pi^2 \sim  N_c$ while the
topological susceptibility $\chi_t^0$ of the quenched (zero flavor) theory
is of order 1. Hence, as $N_c$ goes to infinity, the $\eta'$-meson becomes a 
massless Goldstone boson. The occurrence of the $\eta'$ field in the chiral
Lagrangian has a dramatic effect on the $\theta$-dependence. Since $\theta$
enters as the determinant of the mass matrix, it can be absorbed into a 
redefinition of the field $\eta'$ such that
\begin{eqnarray}
\label{Chact3}
S[\tilde U]&=&\int d^4x \ \{\frac{F_\pi^2}{4} 
\mbox{Tr}[\partial_\mu \tilde U^\dagger \partial_\mu \tilde U] -
\frac{\langle \Psibar \Psi \rangle}{2 N_f} 
\mbox{Tr}[{\cal M} \tilde U^\dagger + {\cal M}^\dagger \tilde U]\} \nonumber \\
&+&\frac{1}{2} \chi_t^0 (i \log \mbox{det} \tilde U - \theta)^2.
\end{eqnarray}
At large $N_c$ the last term in eq.(\ref{Chact3}) can be neglected and thus all
$\theta$-dependence disappears from the theory even at non-zero quark
masses. This is consistent because the anomaly disappears in the large $N_c$ 
limit and $U_A(1)$ chiral rotations can be used to relate different
$\theta$-vacua which thus become physically indistinguishable.

At large but finite $N_c$ we can use the action of eq.(\ref{Chact3}) to derive
the $\theta$-dependence of the Goldstone boson masses for $N_f$ degenerate
quark flavors of mass $m$. First, we must find the minimum of the action, which
turns out to be at
\begin{equation} 
\tilde U = \exp(i \sqrt{2/N_f} \eta'_0(\theta)/F_\pi) \1,
\end{equation}
where
\begin{equation}
\label{eta0}
\langle \Psibar \Psi \rangle m \sin(\sqrt{\frac{2}{N_f}} 
\frac{\eta'_0(\theta)}{F_\pi})
+ \chi_t^0 N_f (\sqrt{2 N_f} \frac{\eta'_0(\theta)}{F_\pi} - \theta) = 0.
\end{equation}
When one is closer to the chiral than to the large $N_c$ limit the second term
on the left-hand side of eq.(\ref{eta0}) dominates and then
\begin{equation}
\eta'_0 = \frac{F_\pi}{\sqrt{2 N_f}} \theta.
\end{equation}
It is easy to show that the vacuum energy takes the form
\begin{equation}
e_0(\theta) = - m \langle\Psibar \Psi\rangle
\cos(\sqrt{\frac{2}{N_f}} \frac{\eta'_0(\theta)}{F_\pi}) +
\frac{1}{2} \chi_t^0 (\sqrt{2 N_f} 
\frac{\eta'_0(\theta)}{F_\pi} - \theta)^2.
\end{equation}
The $\theta$-dependence of the pion mass
\begin{equation}
\label{piontheta}
M^2_\pi(\theta) = M_\pi^2(0)
\cos(\sqrt{\frac{2}{N_f}} \frac{\eta'_0(\theta)}{F_\pi}),
\end{equation}
is similar to the one in eq.(\ref{thetamass}) except that $\theta$ is now 
replaced by $\sqrt{2 N_f} \eta'_0(\theta)/F_\pi$. The $\eta'$ mass is given by
\begin{equation}
\label{etatheta}
M^2_{\eta'}(\theta) = \frac{2 N_f \chi_t^0}{F_\pi^2} + M_\pi^2(\theta).
\end{equation}
At $N_c = \infty$ the first term on the right-hand side of eq.(\ref{etatheta})
vanishes, thus making $\eta'$ degenerate with the pion. Solving explicitly for 
$\eta'_0$, we also note that in the limits $m \rightarrow 0$ or 
$N_c \rightarrow \infty$ physical quantities become $\theta$-independent.

Recently, the dependence of $M_\pi$ and $M_{\eta'}$ on $Q$ has been studied in 
lattice calculations \cite{Bal01}. Each mass was evaluated in two 
domains, one with $|Q| \leq 1$ and the other with $|Q| \geq 2$. While the pion 
masses in both domains agreed with each other within statistical errors of a 
few percent, the $\eta'$-mass was about 15 percent heavier in the  
$|Q| \geq 2$ domain than in the $|Q| \leq 1$ domain. Unfortunately, the quark 
masses used in current lattice calculations are far from the chiral limit such
that the above results from chiral perturbation theory should not be applicable
to these lattice data. For example, in a typical full QCD lattice calculation 
$M_{\pi}/M_{\rho} \ge 0.6$. Still, it is tempting to compare our predictions 
with the lattice data. Assuming that we are closer to the chiral than to the 
large $N_c$ limit, using eqs.(\ref{eta0},\ref{massQ}) we find
\begin{equation}
M_{\pi Q} = M_\pi(0)\left[1 - \frac{1}{4 N_f^2} \frac{1}{\langle Q^2 \rangle}
(1 - \frac{Q^2}{\langle Q^2 \rangle})\right].
\end{equation}
In the lattice calculation of \cite{Bal01} $\beta V = 6.81 \mbox{fm}^4$, 
$\chi_t = 0.70 \mbox{fm}^{-4}$, and hence $\langle Q^2 \rangle = \beta V \chi_t
= 4.75$. This means that the predicted shift of the effective pion mass at
$Q = 0$ is given by $- M_\pi(0)/(4 N_f^2 \langle Q^2\rangle) = 0.013 M_\pi(0)$.
Hence, the effect of fixed topology is of the order of 1 percent, consistent 
with the lattice results.

At large but finite $N_c$ and close enough to the chiral limit (i.e. when
eqs.(\ref{piontheta},\ref{etatheta}) apply) the second $\theta$-derivatives 
$M_{\eta'}''(0)$ and $M_\pi''(0)$ are equal. Hence, using eq.(\ref{massQ}) we 
would expect that the $Q$-dependent shift of the effective masses of $\eta'$ 
and the pion are the same, in disagreement with the lattice results.

\section{$Q$-Dependence in an Instanton Gas}

It is useful to try to understand how fixed topology affects QCD in the domain 
explored by actual lattice calculations. Motivated by the phenomenological 
success of the Veneziano-Witten formula \cite{Ven79,Wit79} relating the 
$\eta'$-mass to the quenched topological susceptibility $\chi_t^0$, in this 
section we will seek a qualitative understanding of the $Q$-dependence of the 
$\eta'$-mass by modeling the fluctuations of the topological charge by an 
instanton gas.

We begin by assuming independent Poisson distributions for instantons and 
anti-instantons in a space-time volume $\beta V$ with 
$\langle N \rangle = \langle \bar N \rangle = \lambda$. Then the probability of
having $N$ instantons and $\bar N$ anti-instantons is given by
\begin{equation}
P(N,\bar N) = \frac{\lambda^{(N + \bar N)}}{N! \bar N!} e^{- 2 \lambda}.
\end{equation}
Using $Q = N - \bar N$ one obtains
\begin{equation}
\label{NNbar}
\langle Q^2 \rangle = \langle N + \bar N \rangle = 2 \lambda = 
\beta V \chi_t^0.
\end{equation}
Fixing the topological charge to $Q$ the probability distribution takes the 
form
\begin{equation}
P_Q(N,\bar N) = \frac{1}{Z_Q} \frac{1}{2 \pi} \int_{-\pi}^\pi d\theta \ 
\frac{\lambda^{N + \bar N}}{N! \bar N!} e^{- 2 \lambda}
\exp[i \theta (Q - (N - \bar N))],
\end{equation}
where
\begin{equation}
Z_Q = \frac{1}{2 \pi} \int_{-\pi}^\pi d\theta \ 
\exp[- 2 \lambda(1 - \cos\theta)] \exp(i \theta Q).
\end{equation}

Motivated by eq.(\ref{NNbar}) we want to model the $Q$-dependence of the 
$\eta'$-mass by assuming
\begin{equation}
\label{etainst}
M_{\eta' Q}^2 = \frac{2 N_f}{F_\pi^2} 
\frac{\langle N + \bar N \rangle_Q}{\beta V}.
\end{equation}
In the large $\beta V$ limit (with $Q/\sqrt{\beta V}$ fixed) one obtains
\begin{equation}
\label{NNbarinst}
\frac{\langle N + \bar N \rangle_Q}{\beta V} = \chi_t^0 - \frac{1}{2 \beta V} 
\left[1 - \frac{Q^2}{\beta V \chi_t^0}\right]. 
\end{equation}
This is a simple and physically appealing result. When $M^2_{\eta' Q}$ is 
averaged with the distribution $Z_Q \sim \exp(- Q^2/2 \langle Q^2 \rangle)$ in 
the appropriate limit, the $\eta'$ mass is consistent with the Veneziano-Witten
formula. The finite volume corrections provide the desired $Q$-dependence.
Using eqs.~(\ref{etainst},\ref{NNbarinst}) for $M_{\eta'}^2$, we find that 
the percentage shift of $M_{\eta'}$ at $Q=0$ is given by 
$1/4\beta V \chi_t^0 = 1/4\langle Q^2\rangle$. In the lattice calculation 
of \cite{Bal01} $1/4 \langle Q^2 \rangle \approx 0.06$, which is of the same 
order of magnitude as the observed $10$ percent shift in the $\eta'$-mass 
between the sum over all $Q$ and the sum over $|Q| < 1.5$. 

\section{Conclusions}

In this work, we have shown how to  correct finite size effects that occur in 
lattice QCD calculations at fixed topology. We have derived a formula that
relates the $Q$-dependent effective mass $M_Q$ to the true mass $M(0)$ and its 
second derivative $M''(0)$ at vacuum angle $\theta = 0$. The difference
$M_Q - M(0)$ is of the order $1/\beta V$ and thus vanishes quickly in the
large volume limit. By observing the predicted finite size effect in a lattice
calculation that gets stuck in fixed topological charge sector, one can 
identify not only $M(0)$ but also $M''(0)$ and thus one can learn something
about the $\theta$-dependence. Close to the chiral limit, we have used chiral
perturbation theory to predict $M''(0)$ for the pion and the nucleon. For large
$N_c$ we have also derived this quantity for the $\eta'$-meson. Our formulae
are in reasonable agreement with existing lattice data for the effective pion 
mass at fixed $Q$, despite the fact that those calculations are performed far
from the chiral limit. On the other hand, perhaps not surprisingly, the 
observed $Q$-dependent effective $\eta'$-mass does not agree with the large 
$N_c$ chiral prediction. Interestingly, a dilute instanton gas model is
qualitatively consistent with the lattice data.

\section*{Acknowledgments}

This work was supported in part by funds provided by the U.S. Department of
Energy (D.O.E.) under cooperative research agreements DE-FC02-94ER40818,
DOE-FG02-91ER40676 and DE-FG02-96ER40945, by the Schweizerischer 
Nationalfond (SNF), and by the European Community's Human Potential 
Programme under contract HPRN-CT-2000-00145.

\end{document}